\documentclass{aa}            
\usepackage{graphicx}
\usepackage{amsmath,amssymb}
\usepackage{multirow,subcaption,float,braket}
\usepackage{booktabs}
\usepackage{makecell}
\usepackage{siunitx}
\usepackage{txfonts}           
\usepackage{hyperref}
\hypersetup{hidelinks}

\begin{document}

  \title{Depolarization and Polarization-Transfer Rates for Solar
     He\,\textsc{i} Lines due to Collisions with Neutral Hydrogen}

  \author{M. Derouich
    \and
    S. Qutub
  }

  \institute{Astronomy Department, Faculty of Science,
       King Abdulaziz University, 21589 Jeddah, Saudi Arabia\\
       \email{aldarwish@kau.edu.sa; squtub@kau.edu.sa}}

  \date{Accepted for publication}

\abstract
{Neutral helium (He\,\textsc{i}) produces several spectral lines that are
widely used for solar diagnostics. The role of collisions between
He\,\textsc{i} atoms and neutral hydrogen (H\,\textsc{i}) in the modeling
of solar He\,\textsc{i} lines remains insufficiently quantified.
Accurate determination of collisional rates affecting atomic polarization
is needed for solar spectropolarimetry.}
{
Our aim is to provide  a set of
multi-level and multi-term collisional depolarization,
polarization-transfer, and population-transfer rates, due to isotropic
collisions with neutral hydrogen, for He\,\textsc{i} levels and terms
involved in the main solar He\,\textsc{i} diagnostic lines.}
{The calculations are performed within the frozen-core approximation, in
which the inner \(1s\) electron is treated as a core with $L_c=0$,
$S_c=1/2$, and \(J_c=1/2\), while the outer electron is treated as the
active valence electron.}
{We compute both multi-level rates, describing depolarization and
polarization transfer between fine-structure $\mathcal{J}$-levels, and
multi-term rates, which additionally account for coherences between
different $\mathcal{J}$-levels belonging to the same term.}
{Our results provide the collisional input needed for the
statistical equilibrium equations (SEE) of the polarization of the main
He\,\textsc{i} solar lines, including the \(10830\,\text{\AA}\),
D$_3$ \(5876\,\text{\AA}\), and related triplet transitions, and allow a
quantitative reassessment of the role of neutral-hydrogen collisions in
He\,\textsc{i} spectropolarimetry.}


\authorrunning{Derouich \& Qutub}
\titlerunning{Depolarization of the solar He\,\textsc{i} lines}

\maketitle

%
%

\section{Introduction}
\label{intro}
The magnetic field is the main driver of activity in the upper solar atmosphere, but its quantitative determination above the photosphere remains a major challenge in solar physics (see, e.g., Wiegelmann, Thalmann \& Solanki 2014; de la Cruz Rodr\'iguez \& Leenaarts 2024). In the chromosphere, prominences, spicules, filaments, and coronal structures, the interpretation of polarized spectral lines generally requires a quantitative description of several atomic processes, including anisotropic radiation pumping, atomic polarization, Hanle effect, and collisional effects (e.g.,  Hanle 1924, Sahal-Br\'echot 1977; Stenflo 1994; Landi Degl'Innocenti \& Landolfi 2004; Derouich 2020).

Among the most important diagnostics of the outer solar atmosphere are the neutral helium multiplets at $10830\,\text{\AA}$ ($2s\,{}^3S \rightarrow 2p\,{}^3P^\circ$) and at $5876\,\text{\AA}$, the D$_3$ line ($2p\,{}^3P^\circ \rightarrow 3d\,{}^3D$). These lines are observed both on the disk and off the limb, and their polarization is sensitive to magnetic fields from fractions of a gauss to several hundred gauss, a range highly relevant for prominences, filaments, spicules, active-region chromospheres, and eruptive structures (e.g., Trujillo Bueno et al. 2002; Casini et al. 2003; Asensio Ramos, Trujillo Bueno, \& Landi Degl'Innocenti 2008; Trujillo Bueno \& del Pino Alem\'an 2022). The metastable lower term of the He\,\textsc{i} $10830\,\text{\AA}$ multiplet makes this line especially useful for probing the upper chromosphere, while the D$_3$ line provides a complementary diagnostic in off-limb plasma.

 Recent observations further illustrate the diagnostic relevance of these helium lines. High-sensitivity He\,\textsc{i} D$_3$ spectropolarimetry with ZIMPOL-3 at IRSOL has been used to infer the magnetic and thermodynamic structure of an active-region prominence (e.g., Esteban Pozuelo et al. 2025), while imaging spectropolarimetry in the He\,\textsc{i} $10830\,\text{\AA}$ triplet has revealed high-speed flows and transition-region-like temperatures during flux emergence (e.g., Leenaarts et al. 2025). Present facilities such as DKIST, together with planned next-generation facilities such as EST, will provide increasingly sensitive visible and near-infrared spectropolarimetry of the chromosphere and low corona (e.g., Rimmele et al. 2020; Quintero Noda et al. 2022), making reliable collisional rates for He\,\textsc{i} polarization modeling increasingly important.

 The development of quantum-mechanical modeling, polarized radiative-transfer
calculations, and inversion techniques has progressively established the
He\,\textsc{i} D$_3$ and \(10830\,\text{\AA}\) multiplets as standard tools
for diagnosing magnetic fields in the solar atmosphere. Early theoretical
studies of scattering polarization and the Hanle effect in the
He\,\textsc{i} D$_3$ line demonstrated its diagnostic potential for
prominence magnetic fields (e.g. Bommier \& Sahal-Br\'echot 1978;
Landi Degl'Innocenti 1982). This diagnostic framework was subsequently
extended and applied to a wide variety of solar structures, including
prominences and filaments, spicules, and active-region chromospheres
(e.g. L\'opez Ariste \& Casini 2002; Trujillo Bueno et al. 2002, 2005;
Casini et al. 2003; Merenda et al. 2006; Kuckein et al. 2009, 2020;
Centeno et al. 2010; Sasso et al. 2011; Orozco Su\'arez et al. 2014;
Schad et al. 2016; D\'iaz Baso et al. 2019; Anan et al. 2021).
However, one physical ingredient remains insufficiently constrained:
the role of elastic collisions with neutral hydrogen. Vicente Ar\'evalo
et al. (2023) pointed out that such collisions are usually assumed to
produce only weak depolarization of He\,\textsc{i} under typical
chromospheric and prominence conditions, but also emphasized that this
assumption has not been investigated in detail and may need to be tested
in sufficiently dense active-region filaments. 

In polarized non-LTE modeling, isotropic collisions with neutral hydrogen enter the statistical-equilibrium equations (SEE) together with radiative and magnetic terms. They can relax the multipole moments of the atomic density matrix, including the alignment responsible for linear polarization, and can transfer population and polarization between fine-structure levels or between coherences. Their importance depends on the local plasma conditions and on their competition with radiative and magnetic processes. Therefore, even when collisions are expected to be weak in many prominence or chromospheric conditions, quantitative rates are required in order to assess this assumption rather than impose it a priori.

A general theoretical framework for collisions with neutral hydrogen has been developed over several decades. The Anstee--Barklem--O'Mara (ABO) theory of collisional line broadening (e.g. Anstee \& O'Mara 1991, 1995; Barklem \& O'Mara 1997; Barklem et al. 1998) was extended to the depolarization and polarization-transfer problem by Derouich, Sahal-Br\'echot, Barklem (DSB), and collaborators (e.g. Derouich et al. 2003a,b, 2004a,b; Derouich et al. 2005b; Derouich et al. 2006; Sahal-Br\'echot et al. 2007; Derouich 2020).  Derouich (2020) summarized extensive calculations for simple atoms in $p$-, $d$- and $f$-states into analytical variation laws as functions of the effective principal quantum number $n^{*}$, while Derouich et al. (2005b) treated $s$-states.

In the present work, we adopt a frozen-core approximation in which the inner $1s$ electron is treated as a core with total angular momentum $J_c=1/2$, while the outer electron is treated as the active valence electron. The collision with neutral hydrogen is assumed to act mainly on this valence electron, with the core angular momentum conserved during the collision. This makes it possible to construct He\,\textsc{i} collisional rates from simple-atom rates by means of angular-momentum recoupling relations involving Wigner $9j$ symbols. A similar approach was also applied when calculating hyperfine structure rates via calculation of fine-structure rates (see, e.g., Nienhuis 1976; Omont 1977; Derouich 2020), where a frozen-nuclear-spin approximation is adopted.

The adopted helium atomic model includes the low-lying singlet and triplet terms relevant for the main solar He\,\textsc{i} diagnostics: 
$1s2s\,{}^3S$, $1s2p\,{}^3P$, $1s2p\,{}^1P $, 
$1s3s\,{}^3S$, $1s3p\,{}^3P $, $1s3p\,{}^1P $, 
$1s3d\,{}^3D$, and $1s3d\,{}^1D$, with the corresponding fine-structure levels resolved where applicable. 
The singlet and triplet systems are treated within the same model to provide a more complete atomic description and to facilitate future extensions that include inelastic processes, such as electron-impact excitation, capable of coupling the two systems.

The aim of this paper is to provide collisional depolarization, polarization-transfer, and population-transfer rates for He\,\textsc{i} levels due to isotropic collisions with neutral hydrogen. The paper is organized as follows. Section~2 describes the effect of isotropic collisions on the atomic density matrix and introduces the coupling scheme used to construct the He\,\textsc{i} rates. Section~3 presents the extension from the multi-level to the multi-term formulation. Section~4 explains how the rates are inferred from simple-atom variation laws and gives the resulting tables. Section~5 summarizes the main conclusions.
\section{Effect of isotropic collisions on the atomic polarization}

\subsection{Density-matrix elements and collisional contribution}

In the density-matrix formalism (e.g. Blum 1981; Landi Degl'Innocenti \& Landolfi 2004), the excitation state of an atomic level
$(\alpha \, \mathcal{J})$ is described by the statistical tensors 
$\rho^{k_{\mathcal J}}_{q}(\alpha \, \mathcal{J})$.
The multipole of rank $k_\mathcal{J}=0$ represents the population of the level,
odd ranks, in particular $k_\mathcal{J}=1$, describe orientation, and even ranks
with $k_\mathcal{J}>0$, in particular $k_\mathcal{J}=2$, describe alignment-type atomic polarization. The  linear polarization is directly related to the alignment components,
while the circular polarization is related to the orientation components (e.g. Landi Degl'Innocenti \& Landolfi 2004).
In the case where the emitting atom is excited by anisotropic but unpolarized radiation, only even values of $k_{\mathcal J}$ are directly created, following the usual optical-pumping selection rules, implying that only linear polarization is observed.

Isotropic collisions with neutral hydrogen modify the atomic polarization
because they tend to equalize the populations of the Zeeman sublevels and
to destroy the coherences between them (e.g. Derouich et al. 2003a,b). In other words, collisions act
directly on the density-matrix elements and, as a consequence, on the
polarization of the emergent radiation. Under the impact approximation,
collisions are assumed to be binary, complete, and well separated in time,
so that the collisional rates are proportional to the hydrogen density
$n_{\mathrm H}$.

In a multi-level atom, the SEE are obtained by
including all the relevant processes that intervene during line formation.
The relative importance of these processes depends on the physical conditions
of the medium where the line is formed. In solar applications, one generally
has to take into account radiative, magnetic, and collisional contributions.
In the stationary regime, one has
\begin{equation}
\left(\frac{d}{dt}\rho^{k_{\mathcal J}}_{q}(\alpha \mathcal{J})\right)_{\mathrm{rad}}
+
\left(\frac{d}{dt}\rho^{k_{\mathcal J}}_{q}(\alpha \mathcal{J})\right)_{\mathrm{mag}}
+
\left(\frac{d}{dt}\rho^{k_{\mathcal J}}_{q}(\alpha \mathcal{J})\right)_{\mathrm{coll}}
=0.
\label{eq:SEE_general}
\end{equation}

If collisions are neglected, or if the corresponding
collisional rates are poorly known, part of the information encoded in the
spectropolarimetric observations may be lost or misinterpreted. In particular,
since the magnetic field is often one of the main unknowns of the SEE, uncertainties in the collisional terms can propagate
directly into the inferred magnetic field and lead to an inaccurate
determination of its strength and orientation.

For a level $(\alpha \, \mathcal{J})$, the collisional contribution to the SEE can be written as (see, e.g., Sahal-Br\'echot et al. 2007; Derouich 2020):
\begin{equation}
\begin{split}
&\left(\frac{d}{dt}\rho^{k_{\mathcal J}}_{q}(\alpha \mathcal{J})\right)_{\mathrm{coll}}
= - \rho^{k_{\mathcal J}}_{q}(\alpha \mathcal{J})\,D^{k_{\mathcal J}}(\alpha \mathcal J) \\
&\quad - \rho^{k_{\mathcal J}}_{q}(\alpha \mathcal{J})
\sum_{\mathcal{J}' \ne \mathcal{J}}
\sqrt{\frac{2\mathcal{J}'+1}{2\mathcal{J}+1}}\,D^{0}(\alpha \mathcal{J} \to \alpha \mathcal{J}') \\
&\quad
+
\sum_{\mathcal{J}' \ne \mathcal{J}}
D^{k_{\mathcal J}}(\alpha \mathcal{J}' \to \alpha \mathcal{J})\,
\rho^{k_{\mathcal J}}_{q}(\alpha \mathcal{J}')
\end{split}
\label{eq:SEE_coll}
\end{equation}
The first term on the right-hand side represents the relaxation of the
multipole $\rho^{k_{\mathcal J}}_{q}(\alpha \mathcal{J})$ by collisions, whereas the second one
describes the transfer of polarization from  $(\alpha \mathcal{J}')$ toward the level $(\alpha \mathcal{J})$. Since the collisions are
isotropic, all the components $q$ of a given tensorial rank $k_{\mathcal J}$ are affected
in the same way.

The quantity $D^{k_{\mathcal J}}(\alpha \mathcal J)$ is the collisional depolarization
 rate of rank $k_{\mathcal J}$ for the level $(\alpha \mathcal J)$. For purely elastic collisions, the population of the level is 
 conserved, so that $D^{0}(\alpha \mathcal J)=0$. Note that, for $k_{\mathcal J}=0$, one has the population-transfer rate $D^{0}(\alpha \mathcal{J} \to \alpha \mathcal{J}')$. The polarization-transfer rate of rank $k_{\mathcal J}$ from $(\alpha\mathcal{J})$ to $(\alpha\mathcal{J}')$ is denoted 
by $D^{k_{\mathcal J}}(\alpha \mathcal{J} \to \alpha \mathcal{J}')$. The \emph{total} rank-$k_{\mathcal J}$ relaxation rate of the level $(\alpha \mathcal{J})$ is the sum of $D^{k_{\mathcal J}}(\alpha \mathcal J)$ and the transfer rate contribution $\sum_{\mathcal{J}' \ne \mathcal{J}}\sqrt{(2\mathcal{J}'+1)/(2\mathcal{J}+1)}\,D^{0}(\alpha \mathcal{J} \to \alpha \mathcal{J}')$.

 \subsection{Coupling scheme }
\label{subsec:atomic_model}

The He\,\textsc{i} atom has two electrons. For the singly excited states
considered in this work, one electron remains in the \(1s\) shell, while
the second electron occupies an \(nl\) orbital. We use the frozen-core
approximation, in which the inner \(1s\) electron is treated as a core and
the outer electron is treated as the active valence electron.
 Within the frozen-core approximation (see, e.g., Derouich et al. 2005a;  Derouich 2020), singly excited He\, \textsc{i}  is described by assuming that one electron remains in the \(1s\) shell and forms the frozen core, while the second electron in the \(nl\) orbital is treated as the active valence electron. The collision with neutral hydrogen is assumed to affect only this outer electron, whose orbital and spin angular momenta are \(l\) and \(s=1/2\), respectively. As a consequence, the core angular momentum is conserved during the collision, and the depolarization and polarization-transfer rates for He\, \textsc{i} may be expressed as linear combinations of the corresponding rates for a simple atom associated with the external \(nl\) electron. The relevant coupling scheme is defined by the valence-electron angular momenta \(l\) and \(s\), its total angular momentum \(j=l+s\), the core angular momenta \(L_c\) and \(S_c\), the core total angular momentum \(J_c=L_c+S_c\), and the total atomic angular momentum \(\mathcal{J}=J_c+j\). For all He\, \textsc{i} levels considered here, the core corresponds to the \(1s\,{}^{2}S_{1/2}\) electron, so that \(L_c=0\), \(S_c=1/2\), and \(J_c=1/2\).

 Such levels can be treated with Eqs.~(\ref{eq:depol}) and
(\ref{eq:transfer}), where \((\alpha \mathcal{J})\) denotes the He\,\textsc{i} level under
consideration. In this picture, the inner \(1s\) electron defines the core, whereas
the outer electron is treated as the valence, or optical, electron.
We adopt a He\, \textsc{i} model atom including the lowest singlet and
triplet terms arising from the configurations \(1s^{2}\), \(1s\,2s\), \(1s\,2p\),
\(1s\,3s\), \(1s\,3p\), and \(1s\,3d\), with the relevant fine-structure
\(\mathcal{J}\)-levels resolved for polarization and magnetic diagnostics. The model
comprises the ground state \(1s^{2}\; {}^{1}S_{0}\); the \(n=2\) terms
\(1s\,2s\, \;{}^{3}S_{1}\), \(1s\,2s\,\;{}^{1}S_{0}\),
\(1s\,2p\,\;{}^{3}P_{\mathcal{J}}\) (\(\mathcal{J}=0,1,2\)), and
\(1s\,2p\,{}^{1}P_{1}\); and the \(n=3\) terms
\(1s\,3s\,{}^{3}S_{1}\), \(1s\,3s\,{}^{1}S_{0}\),
\(1s\,3p\,{}^{3}P_{\mathcal{J}}\) (\(\mathcal{J}=0,1,2\)),
\(1s\,3p\,{}^{1}P_{1}\),
\(1s\,3d\,{}^{3}D_{\mathcal{J}}\) (\(\mathcal{J}=1,2,3\)), and
\(1s\,3d\,{}^{1}D_{2}\). 
This term structure contains the triplet manifold responsible for the principal
He\, \textsc{i} diagnostic lines, in particular the 10830\,\AA\ transition
\(\bigl(2s\,{}^{3}S_{1}\rightarrow 2p\,{}^{3}P_{\mathcal{J}}\bigr)\), the
D\(_3\) 5876\,\AA\ transition
\(\bigl(2p\,{}^{3}P_{\mathcal{J}}\rightarrow 3d\,{}^{3}D_{\mathcal{J}}\bigr)\),
as well as the 7065\,\AA\ and 3889\,\AA\ lines. Under chromospheric and prominence
conditions, the long-lived metastable \(2s\,{}^{3}S_{1}\) level efficiently populates
the triplet ladder, making this model atom a standard framework for describing the
polarization of the strongest He\,\textsc{i} triplet lines in solar applications.

The complete set of quantum numbers used in the multi-level formulation is summarized in Table~\ref{Table_model}. This table specifies, for each configuration and term included in the adopted He\,\textsc{i} atomic model, the values of \(L\), \(S\), the fine-structure levels \(\mathcal{J}'\) and \(\mathcal{J}\), the valence-electron orbital angular momentum \(l\), the contributing valence-electron channels \(j'\) (final level) and \(j\) (initial level), and the allowed even tensorial ranks \(k_\mathcal{J}\). It therefore provides the angular-momentum basis required for applying the frozen-core recoupling relations in Eqs.~(\ref{eq:depol}) and (\ref{eq:transfer}) and for constructing the collisional depolarization and polarization-transfer rates used throughout this work. For depolarization rates, the two levels coincide ($\mathcal{J}=\mathcal{J}'$, $j'=j$). For transfer rates ($\mathcal{J}\neq\mathcal{J}'$), the table lists only the off-diagonal channels with $j\neq j'$, a diagonal channel $j=j'$ does not contribute to a complex-atom transfer rate. 

\begin{table}[htbp]
\centering
\renewcommand{\arraystretch}{1.1}
\caption{He\,\textsc{i} atomic model adopted in the multi-level formulation of this work. The columns list the configuration, term, $L$, $S$, the considered fine-structure levels $\mathcal{J}'$ and $\mathcal{J}$, the orbital quantum number $l$ of the valence electron, the contributing valence-electron channels $j'$ (final level) and $j$ (initial level), and the allowed even tensorial ranks $k_{\mathcal J}$. For depolarization rates $\mathcal{J}=\mathcal{J}'$ and $j'=j$. For transfer rates ($\mathcal{J}\neq\mathcal{J}'$) only off-diagonal channels with $j\neq j'$ are considered.}
\label{Table_model} 
\resizebox{0.95 \columnwidth}{!}{%
\begin{tabular}{llccccc@{\hspace{8pt}}c@{\hspace{8pt}}cc}
\hline
Configuration & Term & $L$ & $S$ & $\mathcal J$ & $\mathcal J'$ & $l$ & $j$ & $j'$ & $k_{\mathcal J}$ \\ [4pt]
\hline 

1s2s & $^3S$ & 0 & 1 & 1 & 1 & 0 & $\frac{1}{2}$ & $\frac{1}{2}$ & 2 \\ [4pt]

1s2p & $^3P$ & 1 & 1 & 2 & 2 & 1 & $\frac{3}{2}$ & $\frac{3}{2}$ & $2,4$ \\ [4pt]
 & & 1 & 1 & 1 & 1 & 1 & $\frac{1}{2}, \, \frac{3}{2}$ & $\frac{1}{2}, \, \frac{3}{2}$ & 2 \\ [4pt]
 & & 1 & 1 & 1 & 2 & 1 & $\frac{1}{2}$ & $\frac{3}{2}$ & $0,2$ \\ [4pt]
 & & 1 & 1 & 0 & 2 & 1 & $\frac{1}{2}$ & $\frac{3}{2}$ & $0$ \\ [4pt]
 & & 1 & 1 & 0 & 1 & 1 & $\frac{1}{2}$ & $\frac{3}{2}$ & $0$ \\ [4pt]

1s2p & $^1P$ & 1 & 0 & 1 & 1 & 1 & $\frac{1}{2}, \, \frac{3}{2}$ & $\frac{1}{2}, \, \frac{3}{2}$ & 2 \\ [4pt]

1s3s & $^3S$ & 0 & 1 & 1 & 1 & 0 & $\frac{1}{2}$ & $\frac{1}{2}$ & 2 \\ [4pt]

1s3p & $^3P$ & 1 & 1 & 2 & 2 & 1 & $\frac{3}{2}$ & $\frac{3}{2}$ & $2,4$ \\ [4pt]
 & & 1 & 1 & 1 & 1 & 1 & $\frac{1}{2}, \, \frac{3}{2}$ & $\frac{1}{2}, \, \frac{3}{2}$ & 2 \\ [4pt]
 & & 1 & 1 & 1 & 2 & 1 & $\frac{1}{2}$ & $\frac{3}{2}$ & $0,2$ \\ [4pt]
 & & 1 & 1 & 0 & 2 & 1 & $\frac{1}{2}$ & $\frac{3}{2}$ & $0$ \\ [4pt]
 & & 1 & 1 & 0 & 1 & 1 & $\frac{1}{2}$ & $\frac{3}{2}$ & $0$ \\ [4pt]

1s3d & $^3D$ & 2 & 1 & 3 & 3 & 2 & $\frac{5}{2}$ & $\frac{5}{2}$ & $2,4,6$ \\ [4pt]
 & & 2 & 1 & 2 & 2 & 2 & $\frac{3}{2}, \, \frac{5}{2}$ & $\frac{3}{2}, \, \frac{5}{2}$ & $2,4$ \\ [4pt]
 & & 2 & 1 & 1 & 1 & 2 & $\frac{3}{2}$ & $\frac{3}{2}$ & 2 \\ [4pt]
 & & 2 & 1 & 2 & 3 & 2 & $\frac{3}{2}$ & $\frac{5}{2}$ & $0,2,4$ \\ [4pt]
 & & 2 & 1 & 1 & 3 & 2 & $\frac{3}{2}$ & $\frac{5}{2}$ & $0,2$ \\ [4pt]
 & & 2 & 1 & 1 & 2 & 2 & $\frac{3}{2}$ & $\frac{5}{2}$ & $0,2$ \\ [4pt]

1s3d & $^1D$ & 2 & 0 & 2 & 2 & 2 & $\frac{3}{2}, \, \frac{5}{2}$ & $\frac{3}{2}, \, \frac{5}{2}$ & $2,4$ \\ [4pt]

1s3p & $^1P$ & 1 & 0 & 1 & 1 & 1 & $\frac{1}{2}, \, \frac{3}{2}$ & $\frac{1}{2}, \, \frac{3}{2}$ & 2 \\ [4pt]
\hline
\end{tabular}
}
\end{table}

The singlet terms included in this work, together with their depolarization and polarization-transfer rates, are also calculated.
Indeed, when inelastic collisions are included---especially electron-impact excitation---the assumption of a purely triplet model may become overly
restrictive. Although singlet--triplet radiative transitions are forbidden, singlet terms
can still influence the triplet terms populations and polarizations through collisional
coupling (population transfer and polarization transfer) between the two spin systems. For
this reason, we extend the commonly adopted triplet scheme by explicitly including additional
low-lying singlet terms alongside the corresponding triplet terms. This yields a more flexible
and physically complete multi-term atomic model whenever inelastic electron collisions are
accounted for, and it reduces the risk of bias in the inferred polarization signals when
singlet--triplet collisional channels are non-negligible under chromospheric conditions.

By  assuming that the core of the He\,\textsc{i} atom is frozen,  the collisional depolarization rate of a level $(\alpha \mathcal{J})$ of a complex atom can be 
written, following Derouich (2020), as:
\begin{equation}
\begin{aligned}
D^{k_{\mathcal J}}(\alpha \mathcal J , j)
={}& (2\mathcal J+1)^2
\sum_{k_j}(2k_j+1)\,D^{k_j}(j) \\
&\times \sum_{k_{J_c}}(2k_{J_c}+1)
\begin{Bmatrix}
j  & J_c & \mathcal J \\
j  & J_c & \mathcal J \\
k_j & k_{J_c} & k_{\mathcal J}
\end{Bmatrix}^{2},
\end{aligned}
\label{eq:depol} 
\end{equation}
where $J_c$ is the total angular momentum of the core, $j$ is the total angular momentum of the external 
valence electron, and $D^{k_j}(j)$ is the depolarization rate of the external shell treated as that of
a simple atom. The quantities $k_j$, $k_{J_c}$, $k_\mathcal{J}$ are tensorial ranks.  
When two $j$-channels contribute additively to the same level $(\alpha \mathcal J)$ (as can be seen in Table \ref{Table_model}), the 
depolarization rate is obtained by summing the corresponding contributions:
\begin{equation}
D^{k_{\mathcal J}}(\alpha \mathcal J)
=
\sum_{j} D^{k_\mathcal{J}}\!\left( \alpha \mathcal{J} ,j \right) .
\label{eq:depol_total}
\end{equation}
Similarly, the  polarization-transfer rates between fine-structure levels $(\alpha \mathcal{J})$ and $(\alpha \mathcal{J}')$ are given by (see, e.g., Derouich 2020):
\begin{equation}
\begin{aligned}
&D^{k_{\mathcal{J}}}(\alpha \mathcal{J} \to \alpha \mathcal{J}' ,j , j')
=
(2 \mathcal{J}+1)(2 \mathcal{J}'+1) \\
&\quad\times
\sum_{k_j} (2k_j+1)\, D^{k_j}(j \to j')
\sum_{k_{J_c}} (2k_{J_c}+1) \\
&\quad\times
\begin{Bmatrix}
j  & J_c   & \mathcal{J} \\
j  & J_c   & \mathcal{J} \\
k_j & k_{J_c} & k_{\mathcal{J}}
\end{Bmatrix}
\begin{Bmatrix}
j' & J_c   & \mathcal{J}' \\
j' & J_c   & \mathcal{J}' \\
k_j & k_{J_c} & k_{\mathcal{J}}
\end{Bmatrix}
\end{aligned}
\label{eq:transfer} 
\end{equation}
Summing over all valence-electron channels compatible with the
initial and final He\,\textsc{i} fine-structure levels, with $j\neq j'$
(diagonal channels do not contribute to a transfer), yields the
total multi-level transfer rate:
\begin{equation}
D^{k_\mathcal{J}}\!\left(\alpha \mathcal{J} \rightarrow \alpha \mathcal{J}' \right)
=
\sum_{\substack{(j,j')\\[1pt] j\neq j'}} D^{k_\mathcal{J}}\!\left(\alpha \mathcal{J} \rightarrow \alpha \mathcal{J}' ,j,j'\right) ,
\label{eq:transfer_total_multilevel}
\end{equation}
For \(s\)-states, \(l=0\), and only
$j=\frac{1}{2}$
is possible. 
For \(p\)-states, \(l=1\), and the possible channels are
$j=\frac{1}{2},\frac{3}{2}$.
Since \(J_c=\frac{1}{2}\), these channels give
$j=\frac{1}{2} \Rightarrow \mathcal{J}=0,1$
and
$j=\frac{3}{2} \Rightarrow \mathcal{J}=1,2$.
Thus, the level \(\mathcal{J}=1\) can receive contributions from both
\(j=\frac{1}{2}\) and \(j=\frac{3}{2}\). For \(d\)-states, \(l=2\), and the possible
channels are
$j=\frac{3}{2},\frac{5}{2}$.
They give
$j=\frac{3}{2} \Rightarrow \mathcal{J}=1,2$
and
$j=\frac{5}{2} \Rightarrow \mathcal{J}=2,3$.
Thus, the level \(\mathcal{J}=2\) can receive contributions from both
\(j=\frac{3}{2}\) and \(j=\frac{5}{2}\).

For a given He\,\textsc{i} level, the collision rate is first evaluated for each allowed valence-electron channel $j$ and the partial contributions are then added. Thus, levels admitting a single $j$ channel, such as $s$-states, have a single contribution, whereas the levels $\mathcal{J}=1$ in $p$-states and $\mathcal{J}=2$ in $d$-states require a sum over the two allowed channels. The same rule is used for transfer rates. 

The complex-atom polarization- and population-transfer rate $D^{k_{\mathcal{J}}}(\alpha\mathcal{J}\rightarrow\alpha\mathcal{J}')$ is, by construction, a recoupling of the simple-atom \emph{transfer} rate $D^{k_j}(j\rightarrow j')$ with $j\neq j'$. Therefore, a valence-electron channel  contributes to the complex-atom transfer rate only if the corresponding simple-atom transfer rate $D^{k_j}(j\rightarrow j')$ actually exists. This same constraint
applies to the multi-term transfer rates introduced below. By contrast,
the multi-level depolarization rates
$D^{k_{\mathcal{J}}}(\alpha\mathcal{J})$ and the multi-term depolarization rates
discussed in the next section involve the simple-atom depolarization rate
$D^{k_j}(j)$, corresponding to the diagonal case $j=j'$. 

General laws were obtained in Derouich et al. (2005b) for a valence electron in $s$-states and in Derouich (2020) for a valence electron in $p$- and $d$-states; these laws allow to calculate $D^{k_j}(j)$ and $D^{k_j}(j \rightarrow j')$, which are used to infer the collisional rates
needed for modeling the polarization of He\,\textsc{i} lines. For each fine-structure level involved in the
He\,\textsc{i} atomic model, the effective principal quantum number $n^{*}$ is
first determined from the known atomic energy levels.

\section{Multi-term Formulation and Statistical Equilibrium Equations}

In the multi-level formulation, the atomic polarization is described by
the statistical tensors \(\rho^k_q(\alpha \mathcal{J})\), where each fine-structure
level $(\alpha \mathcal{J})$ is treated separately. A more general treatment of scattering polarization is obtained in the \emph{multi-term}
framework (e.g. Landi Degl'Innocenti \& Landolfi 2004), where one allows for coherences not only between Zeeman
sublevels of a given $\mathcal{J}$-level, but also between different $\mathcal{J}$-levels belonging to the same LS term
$\alpha \equiv (n\,l\,L\,S)$. In this case the atomic polarization is described by spherical statistical tensors
$\rho^{k_{\mathcal J}}_{q}(\alpha \mathcal{J} \mathcal{J}')$ that include both diagonal ($\mathcal{J}=\mathcal{J}'$) and off-diagonal ($\mathcal{J}\neq \mathcal{J}'$) elements.

The diagonal elements with \(\mathcal{J}=\mathcal{J}'\) describe the
population and polarization of a given fine-structure level, while the
off-diagonal elements with \(\mathcal{J}\neq \mathcal{J}'\) describe coherences between
different fine-structure levels of the same term.

The allowed tensorial ranks satisfy
\[
|\mathcal{J}-\mathcal{J}'| \leq k_{\mathcal J} \leq \mathcal{J}+\mathcal{J}' .
\]
In the present work, only even values of \(k\) are retained, since they
are the relevant ranks for population and linear-polarization studies.
The rank \(k_\mathcal{J}=0\) corresponds to population, while \(k_\mathcal{J}=2\) corresponds to
alignment. Higher even ranks, such as \(k_\mathcal{J}=4\) and \(k_\mathcal{J}=6\), are also kept
when allowed, because they may be coupled to the lower ranks through the
SEE.

In the multi-term case, isotropic collisions contribute to the  SEE through
relaxation of $\rho^{k_{\mathcal J}}_{q}(\alpha \mathcal{J} \mathcal{J}')$ and through transfer terms coupling different pairs $(\mathcal{J},\mathcal{J}')$ and
$(\mathcal{J}'',\mathcal{J}''')$ within the same term. The collisional contribution can be written as (e.g. Derouich \& Qutub 2024):
\begin{equation}
\small
\begin{split}
\left(\frac{d}{dt}\rho^{k_{\mathcal J}}_{q}(\alpha J J')\right)_{\rm coll}
=&-
\Bigg[
D^{k_{\mathcal J}}(\alpha J J')
+
\sum_{(J''J''')\neq(JJ')}
\sqrt{\frac{J''+J'''+1}{J+J'+1}}
\\
&\times
D^{0}(\alpha J J' \rightarrow \alpha J''J''')
\Bigg]\rho^{k_{\mathcal J}}_{q}(\alpha J J')
\\
&+
\sum_{(J''J''')\neq(JJ')}
D^{k_{\mathcal J}}(\alpha J''J''' \rightarrow \alpha J J')
\rho^{k_{\mathcal J}}_{q}(\alpha J''J''') .
\end{split}
\label{eq:SEE_multiterm_coll}
\end{equation}
\normalsize
where $D^{k_{\mathcal J}}(\alpha \mathcal{J} \mathcal{J}')$ represents the rank-$k_{\mathcal J}$ collisional relaxation of the coherence $(\mathcal{J},\mathcal{J}')$, and
$D^{k_{\mathcal J}}(\alpha \mathcal{J}\mathcal{J}'\rightarrow \alpha \mathcal{J}'' \mathcal{J}''')$ are the polarization/popu\-lation-transfer rates between pairs of
fine-structure levels within the term.

Direct computation of the full set of multi-term rates $D^{k_{\mathcal J}}(\alpha \mathcal{J} \mathcal{J}')$ and 
$D^{k_\mathcal{J}}\!\left(\alpha \mathcal{J} \mathcal{J}' \rightarrow \alpha \mathcal{J}'' \mathcal{J}'''\right)$ is generally difficult because it would require solving the collision
problem while explicitly retaining $\mathcal{J}\!-\!\mathcal{J}'$ coherences. An efficient approach is to employ the
 frozen-core  approximation like for the multi-level case (see, e.g., Derouich et al. 2005b). In this indirect method, one obtains the multi-term rates as:
\begin{equation}
\begin{aligned}
&D^{k_\mathcal{J}} (\alpha \mathcal{J} \mathcal{J}',j)
=(2\mathcal{J}+1)(2\mathcal{J}'+1) \\
&\quad\times
\sum_{k_j} (2k_j+1)\, D^{k_j}(j)
\sum_{k_{J_c}} (2k_{J_c}+1) \\
&\quad\times
\begin{Bmatrix}
j  & J_c & \mathcal{J} \\
j  & J_c & \mathcal{J}' \\
k_j & k_{J_c} & k_\mathcal{J}
\end{Bmatrix}
\begin{Bmatrix}
j  & J_c & \mathcal{J} \\
j  & J_c & \mathcal{J}' \\
k_j & k_{J_c} & k_\mathcal{J}
\end{Bmatrix} 
\end{aligned}
\label{eq:Depo_multi}
\end{equation}
The total multi-term relaxation rate of the coherence
$(\alpha\mathcal{J}\mathcal{J}')$ is obtained by summing the partial
contributions of Eq.~(\ref{eq:Depo_multi}) over the allowed
valence-electron channels $j$ of the term:
\begin{equation}
D^{k_{\mathcal J}}(\alpha \mathcal{J} \mathcal{J}')
= \sum_{j} D^{k_{\mathcal J}}(\alpha \mathcal{J} \mathcal{J}',j) ,
\label{eq:depol_total_multiterm} 
\end{equation}
and
\begin{align}
&D^{k_\mathcal{J}}\!\left(\alpha \mathcal{J} \mathcal{J}' \rightarrow \alpha \mathcal{J}'' \mathcal{J}''',j,j'\right)
\nonumber\\
&\quad =
\sqrt{(2\mathcal{J}+1)(2\mathcal{J}'+1)(2\mathcal{J}''+1)(2\mathcal{J}'''+1)}
\nonumber\\
&\quad \times
\sum_{k_{j}}(2k_{j}+1)\,
D^{k_{j}}(\alpha j \rightarrow \alpha j')
\sum_{k_{J_c}}(2k_{J_c}+1)
\nonumber\\
&\quad\times
\begin{Bmatrix}
j & J_c & \mathcal{J}\\
j & J_c & \mathcal{J}'\\
k_{j} & k_{J_c} & k_\mathcal{J}
\end{Bmatrix}
\begin{Bmatrix}
j' & J_c & \mathcal{J}''\\
j' & J_c & \mathcal{J}'''\\
k_{j} & k_{J_c} & k_\mathcal{J}
\end{Bmatrix} ,
\label{eq:recoupling_multiterm}
\end{align}
and the total multi-term transfer rate is obtained by summing over the allowed valence-electron channels for each pair, in analogy with Eq.~(\ref{eq:depol_total}) for the multi-level case and Eq.~(\ref{eq:depol_total_multiterm}) for the multi-term relaxation:
\begin{equation}
D^{k_\mathcal{J}}\!\left(\alpha \mathcal{J} \mathcal{J}' \rightarrow \alpha \mathcal{J}'' \mathcal{J}'''\right)
=
\sum_{\substack{(j,j')\\[1pt] j\neq j'}} D^{k_\mathcal{J}}\!\left(\alpha \mathcal{J} \mathcal{J}' \rightarrow \alpha \mathcal{J}'' \mathcal{J}''',j,j'\right) ,
\label{eq:transfer_total_multiterm}
\end{equation}
where the sum runs over all possible $(j,j')$ pairs with $j\neq j'$, i.e. only  simple-atom transfer rates $D^{k_j}(j\rightarrow j')$ may enter the channel sum. 
 One has to compute (or infer) the simple-atom rates $D^{k_{j}}(j)$ and $D^{k_{j}}(j\rightarrow j')$ for the valence
electron in order to generate the full set of $D^{k_\mathcal{J}}\!\left(\alpha \mathcal{J} \mathcal{J}'\right)$ and
$D^{k_\mathcal{J}}\!\left(\alpha \mathcal{J} \mathcal{J}' \rightarrow \alpha \mathcal{J}'' \mathcal{J}'''\right)$ needed in the multi-term SEE.

When the Boltzmann factor $e^{-\Delta E/k_{\rm B}T}$ can be approximated by
unity --- as is the case for the small fine-structure splittings of He\,\textsc{i}
at $T=5000$ K, where $\Delta E/k_{\rm B}T$ ranges from about
$1.3\times 10^{-5}$ to $3.1\times 10^{-4}$ within a given term --- the multi-term
collisional transfer rates obey a balance relation that connects the forward
$(\mathcal{J},\mathcal{J}')\rightarrow(\mathcal{J}'',\mathcal{J}''')$ and reverse
$(\mathcal{J}'',\mathcal{J}''')\rightarrow(\mathcal{J},\mathcal{J}')$ processes
through the effective pair weight $g_{\mathcal{J}\mathcal{J}'}=\mathcal{J}+\mathcal{J}'+1$:
\begin{align}
&D^{k}\!\left(\alpha\mathcal{J}''\mathcal{J}'''\rightarrow\alpha\mathcal{J}\mathcal{J}'\right)
\nonumber\\
&\qquad =\frac{\mathcal{J}+\mathcal{J}'+1}{\mathcal{J}''+\mathcal{J}'''+1}\,
D^{k}\!\left(\alpha\mathcal{J}\mathcal{J}'\rightarrow\alpha\mathcal{J}''\mathcal{J}'''\right).
\label{eq:balance_multiterm}
\end{align}
In the diagonal multi-level limit $\mathcal{J}=\mathcal{J}'$ and
$\mathcal{J}''=\mathcal{J}'''$, the statistical weights reduce to
$\mathcal{J}+\mathcal{J}'+1=2\mathcal{J}+1$ and
$\mathcal{J}''+\mathcal{J}'''+1=2\mathcal{J}''+1$, and
Eq.~(\ref{eq:balance_multiterm}) recovers the usual multi-level detailed-balance
relation for the population-transfer rates between two fine-structure levels:
\begin{equation}
D^{k}\!\left(\alpha\mathcal{J}''\rightarrow\alpha\mathcal{J}\right)
=\frac{2\mathcal{J}+1}{2\mathcal{J}''+1}\,
D^{k}\!\left(\alpha\mathcal{J}\rightarrow\alpha\mathcal{J}''\right),
\label{eq:balance_multilevel}
\end{equation}
neglecting the Boltzmann factor. This 
explains why only the upward (lower-energy to higher-energy) rates are
listed in
Tables~\ref{tab:helium_multilevel_transfer_rates} and
\ref{tab:helium_multiterm_pair_transfer}: the downward rates follow from
Eqs.~(\ref{eq:balance_multiterm}) and (\ref{eq:balance_multilevel}) without
further computation. We emphasize that these balance relations connect
the two physically distinct processes of forward and reverse transfer;
they should not be confused with a mere permutation of the indices
$\mathcal{J}\leftrightarrow\mathcal{J}'$ inside a single coherence pair,
which is not a detailed-balance relation.

\section{Inference of He\,\textsc{i} Collisional Rates  from General Variation Laws}
\subsection{Cases in which the valence electron is in $s$-States}
\label{subsec:s_states}
For $s$-states, the valence electron has $l=0$, and therefore only one orbital sublevel is present. If the electron spin is neglected, isotropic collisions cannot couple different magnetic sublevels, the scattering matrix remains diagonal, and the depolarization rate of a spherically symmetric $j=1/2$ level vanishes (see Derouich et al. 2005b). Depolarization can therefore arise only through spin-dependent effects. Although a $j=1/2$ level cannot carry alignment, its destruction of orientation remains important because it enters the recoupling expressions used to obtain depolarization and population-transfer rates in more complex atoms, and because it is required for interpreting polarization in hyperfine-structured levels.
The appropriate treatment of these collisions is the formalism detailed in Derouich et al. (2005b) (see also Derouich \& Barklem 2007), which includes exchange interactions through symmetry-adapted perturbation theory based on the Murrell--Shaw--Musher--Amos formalism. Thus, the collisional physics of $s$-states differs fundamentally from that of $p$- and $d$-states (see next subsection). In the present work, the depolarization and transfer rates associated with $s$-state contributions are therefore computed using the variation laws of Derouich et al. (2005b).

For a simple atom in an $ns\,$ $^2S_{1/2}$ state ($n$ is the principal quantum number), only the destruction of
orientation (tensorial rank $k_j = 1$) is non-zero, while population
($k_j=0$) and higher-rank depolarization rates vanish. At the reference
temperature $T = 5000\,\mathrm{K}$, the destruction rate of orientation follows a
power-law behavior with the effective principal quantum number $n^{*}$ (see Derouich et al. 2005b):
\begin{equation}
D^{k_j=1}(j=1/2, T=5000\,\mathrm{K}) =
1.0045 \times 10^{-9}\, n_{\mathrm{H}}\, (n^{*})^{2.979},
\label{eq:s_state_5000K}
\end{equation}
where $n_{\mathrm{H}}$ is the neutral hydrogen density in cm$^{-3}$ and the rate
is expressed in s$^{-1}$.

The temperature dependence of the destruction of orientation is also described
by a power law (see Derouich et al. 2005b):
\begin{equation}
D^{k_j=1}(j=1/2, T) =
D^{k_j=1}(j=1/2, T=5000\,\mathrm{K})
\left( \frac{T}{5000} \right)^{0.416},
\label{eq:s_state_T}
\end{equation}
which provides an accurate description of the temperature scaling for solar and
stellar atmospheric conditions.

In the present work, Eqs.~(\ref{eq:s_state_5000K}) and (\ref{eq:s_state_T}) are used to evaluate the simple-atom destruction rate of orientation for He\,\textsc{i} levels whose valence electron occupies an $s$-state. For each level, the effective principal quantum number $n^{*}$ should be determined (see Table~\ref{tab:nstar} and Eq.~\ref{eq:nstar_def} in Section~\ref{subsec:full_workflow}), and the corresponding rate is computed at the required temperature. These simple-atom rates are then recoupled to the full He\,\textsc{i} fine-structure system using the angular-momentum relations described in Section~\ref{subsec:atomic_model} (the frozen-core coupling scheme), via Eqs.~(\ref{eq:depol})--(\ref{eq:transfer_total_multilevel}) for the multi-level rates and Eqs.~(\ref{eq:Depo_multi})--(\ref{eq:transfer_total_multiterm}) for the multi-term rates. 

The accuracy of the He\,\textsc{i} rates associated with $s$-state contributions is governed by two distinct ingredients. 

The first is the underlying   simple-atom   $s$-state rate, namely the destruction rate of orientation computed with the spin-dependent semi-classical formalism of Derouich et al. (2005b). This part of the method explicitly includes exchange interactions through symmetry-adapted perturbation theory and has been quantitatively benchmarked: the calculated rate agrees with the fully quantum result for Na\,\textsc{i} to better than $1\%$ at $T=5000\,\mathrm{K}$, and the extension to the simple ion Ca\,\textsc{ii} differs by only about $4\%$ at the same temperature. Thus, the intrinsic uncertainty of the $s$-state prescription itself is small. 

The second ingredient is the frozen-core reduction used to apply these  simple-atom   rates to He\,\textsc{i}. In this approximation, the compact inner $1s$ electron is kept inert, while the collision with neutral hydrogen is assumed to act predominantly on the outer active electron. Once this approximation is adopted, the recoupling to the He\,\textsc{i} fine-structure levels is purely algebraic and introduces no additional dynamical uncertainty. A direct percentage error for the frozen-core approximation in He\,\textsc{i}+H depolarizing collisions cannot yet be assigned, because no full two-electron benchmark calculation is available. Nevertheless, the present case is physically favorable: the helium core contains only one tightly bound $1s$ electron, unlike complex atoms or ions where the frozen core may contain many electrons or open subshells. Moreover, the broader frozen-core/optical-electron strategy has been used successfully in depolarization and line-broadening calculations for complex systems; in the related ABO broadening theory  (see Anstee
1992; Anstee \& O'Mara 1991, 1995; Anstee et al. 1997;
Barklem 1998; Barklem \& O'Mara 1997; Barklem et al. 1998), comparison with solar line profiles and abundance determinations, including agreement with meteoritic abundances, indicates typical accuracies of order $20\%$ or better. Since the present $s$-state treatment additionally includes the spin and exchange effects absent from the standard broadening validation, the generic $20\%$ value should be viewed as a conservative upper-bound reference rather than as the expected uncertainty for He\,\textsc{i}. We therefore regard the present He\,\textsc{i} rates as physically reliable semi-classical estimates, with the dominant remaining uncertainty arising from the frozen-core assumption rather than from either the spin-exchange $s$-state rates or the angular-momentum recoupling.

\subsection{Cases in which the valence electron is in $p$- and $d$-states}
The collisional depolarization rates $D^{k_\mathcal{J}} (\alpha \mathcal{J}) $ and polarization-transfer rates $D^{k_\mathcal{J}} (\alpha \mathcal{J}\rightarrow \alpha \mathcal{J}') $ required in the
present work are inferred using the general variation laws established in
Derouich (2020), which provide a comprehensive and unified description of
(de)polarizing collisions of simple atoms with neutral hydrogen. In that work, extensive numerical calculations based on
the Derouich--Sahal-Br\'echot (DSB) semiclassical approach were performed for hypothetical simple atoms in
$p$- and $d$-states, covering a wide range of effective principal quantum
numbers $n^{*}$. 

The resulting database of thousands of cross sections was then
condensed into 48 analytical variation laws that express the depolarization and
polarization-transfer rates as power laws of $n^{*}$.

\begin{table}[htbp]
\centering
\scriptsize
\setlength{\tabcolsep}{2pt}
\renewcommand{\arraystretch}{1.05}
\caption{Energies and effective principal quantum numbers of the He~{\sc i}
levels included in the atomic model. The energy values are taken from the NIST database (Kramida et al. 2022).}
\label{tab:nstar}
\resizebox{0.95 \columnwidth}{!}{%
\begin{tabular}{@{}llcccccc@{}}
\hline
Configuration & Term & $L$ & $S$ & $\mathcal{J}$ & $l$ & $E$ (a.u.) & $n^{*}$ \\
\hline
$1s2s$ & ${}^3S$ & 0 & 1 & 1 & 0 & 0.7283573995900 & 1.689283 \\
$1s2s$ & ${}^1S$ & 0 & 0 & 0 & 0 & 0.7576157508600 & 1.850874 \\
$1s2p$ & ${}^3P$ & 1 & 1 & 2 & 1 & 0.7704159832056 & 1.937796 \\
$1s2p$ & ${}^3P$ & 1 & 1 & 1 & 1 & 0.7704163314258 & 1.937799 \\
$1s2p$ & ${}^3P$ & 1 & 1 & 0 & 1 & 0.7704208326986 & 1.937831 \\
$1s2p$ & ${}^1P$ & 1 & 0 & 1 & 1 & 0.7797479525500 & 2.009492 \\
$1s3s$ & ${}^3S$ & 0 & 1 & 1 & 0 & 0.8348882417800 & 2.698141 \\
$1s3s$ & ${}^1S$ & 0 & 0 & 0 & 0 & 0.8423061269900 & 2.856822 \\
$1s3p$ & ${}^3P$ & 1 & 1 & 2 & 1 & 0.8454943433507 & 2.934191 \\
$1s3p$ & ${}^3P$ & 1 & 1 & 1 & 1 & 0.8454944435360 & 2.934194 \\
$1s3p$ & ${}^3P$ & 1 & 1 & 0 & 1 & 0.8454956769020 & 2.934225 \\
$1s3d$ & ${}^3D$ & 2 & 1 & 3 & 2 & 0.8479410236568 & 2.998023 \\
$1s3d$ & ${}^3D$ & 2 & 1 & 2 & 2 & 0.8479410350958 & 2.998023 \\
$1s3d$ & ${}^3D$ & 2 & 1 & 1 & 2 & 0.8479412364771 & 2.998029 \\
$1s3d$ & ${}^1D$ & 2 & 0 & 2 & 2 & 0.8479566086486 & 2.998443 \\
$1s3p$ & ${}^1P$ & 1 & 0 & 1 & 1 & 0.8484322820200 & 3.011349 \\
\hline
\end{tabular}%
}
\end{table}

For a given atomic level characterized by the total angular momentum $j$, the
tensorial order $k_j$, and the effective principal quantum number $n^{*}$, the
rates at the reference temperature $T = 5000\,\mathrm{K}$ are given by
\begin{equation}
D^{k_j}(j, T=5000\,\mathrm{K}) =
n_{\mathrm{H}} \times 10^{-9}\, a^{k_j} (j) \; (n^{*})^{\, b^{k_j}(j)},
\label{eq:pdf_state_5000K}
\end{equation}
\begin{equation}
D^{k_j}(j \rightarrow j', T=5000\,\mathrm{K}) =
n_{\mathrm{H}} \times 10^{-9}\, a^{k_j} (j \rightarrow j') \; (n^{*})^{\, b^{k_j}(j \rightarrow j')},
\label{eq:pdf_state_5000K_transfer}
\end{equation}
where $n_{\mathrm{H}}$ is the hydrogen density in cm$^{-3}$ and the coefficients
$a^{k_j}$ and $b^{k_j}$ are tabulated in Derouich (2020). 

 For each fine-structure level involved in the
He\,\textsc{i} atomic model, the effective principal quantum number $n^{*}$ is
first determined from the known atomic energy levels. The corresponding
depolarization and polarization-transfer rates at $T = 5000\,\mathrm{K}$ are then
directly obtained by applying the above variation laws.

Given the broad thermal range of the solar chromosphere, the temperature dependence of the collisional rates must be taken into account. As discussed in Derouich et al. (2006), the
(de)polarization rates due to collisions with neutral hydrogen scale with
temperature as $T^{0.38}$ for $p$- and $d$-states. Therefore, the rates at an arbitrary temperature $T$
are obtained through
\begin{equation}
D^{k_j}(j, T) =
D^{k_j}(j, T=5000\,\mathrm{K})
\left( \frac{T}{5000} \right)^{0.38},
\label{eq:pdf_state_T}
\end{equation}
\begin{equation}
D^{k_j}(j \rightarrow j', T) =
D^{k_j}(j \rightarrow j', T=5000\,\mathrm{K})
\left( \frac{T}{5000} \right)^{0.38} .
\label{eq:pdf_state_T_transfer}
\end{equation}

This procedure allows one to efficiently and consistently determine all the
collisional depolarization, polarization-transfer, and population-transfer
rates required for the multi-level and multi-term modeling of He\,\textsc{i} 
lines, without performing new quantum or semiclassical collision
calculations (Eqs.~\ref{eq:depol}-- \ref{eq:transfer_total_multilevel} and Eqs.~\ref{eq:Depo_multi}--\ref{eq:transfer_total_multiterm}). Consequently, the He\,\textsc{i} rates benefit
from the generality and robustness of the variation laws established in
Derouich (2020), which were designed to be applicable to any simple
atom interacting with neutral hydrogen under solar conditions.

\subsection{Workflow for Computing He\,\textsc{i} Collisional Rates}
\label{subsec:full_workflow}
The complete procedure for computing a He\,\textsc{i} rate is as follows:
\begin{enumerate}
\item Select the He\,\textsc{i} levels $\mathcal{J}$ and $\mathcal{J}'$ from the table of quantum numbers.
\item Read the corresponding quantum numbers $l$, $s$, $j$ and $L_c$, $S_c$, $J_c$.
\item Read the energy $E_{\text{level}}$ from Table~\ref{tab:nstar} and compute $n^*$ using:
\begin{equation}
n^* = \frac{1}{\sqrt{2(E_{\text{ionization}} - E_{\text{level}})}}
\label{eq:nstar_def}
\end{equation}
where, for the He\,\textsc{i} atom, $E_{\text{ionization}} = 0.9035699$~a.u.\ relative to the ground-state energy.
\item Evaluate the simple-atom rates:
\begin{itemize}
\item Use the Derouich (2020) variation-law equations (Eqs.~\ref{eq:pdf_state_5000K}--\ref{eq:pdf_state_T_transfer}) for $p$- and $d$-states 
\item Use the Derouich et al. (2005b) equations (Eqs.~\ref{eq:s_state_5000K} and \ref{eq:s_state_T}) for $s$-states
\end{itemize}
\item Insert the simple-atom rates into the transfer and depolarization rate equations (Eqs.~\ref{eq:depol}-- \ref{eq:transfer_total_multilevel} for the multi-level formulation, or Eqs.~\ref{eq:Depo_multi}--\ref{eq:transfer_total_multiterm} for the multi-term formulation).
\item Perform the linear combination of $9j$-symbol summations to obtain all rates for the atomic model of He\,\textsc{i}. The transfer rates are computed by retaining only off-diagonal valence-electron channels with $j\neq j'$. The depolarization rates are computed from the corresponding simple-atom depolarization channels and then summed over the allowed $j$ contributions.
\end{enumerate}
This procedure provides a consistent and computationally efficient way to construct the full set of He\,\textsc{i} collisional rates needed for the SEE and polarized radiative-transfer calculations.

The resulting multi-level depolarization rates and multi-level population-/polarization-transfer rates are listed in Tables~\ref{tab:helium_multilevel_depolarization_rates} and \ref{tab:helium_multilevel_transfer_rates}, respectively. The multi-term depolarization coefficients and multi-term transfer rates are listed in Tables~\ref{tab:helium_multiterm_rates} and \ref{tab:helium_multiterm_pair_transfer}, respectively.

\begin{table}[htbp]
\centering
\scriptsize
\renewcommand{\arraystretch}{1.1}
\caption{Computed He\,\textsc{i} non-zero multi-level depolarization rates at
$T=5000\,{\rm K}$. The coefficients $a^k$ are defined through
$D^k(\alpha \mathcal{J})=a^k \times 10^{-9} n_{\rm H}$ s$^{-1}$, where
$n_{\rm H}$ is in cm$^{-3}$. The column preceding the tensorial rank $k_{\mathcal J}$ lists the
valence-electron channel(s) $j$ of the simple-atom depolarization rate
entering the recoupling. When more than one channel contributes, all
contributing channels are listed and their partial contributions are summed.}
\label{tab:helium_multilevel_depolarization_rates}
\resizebox{0.95 \columnwidth}{!}{%
\begin{tabular}{llcccc}
\hline
Configuration & Term & $\mathcal{J}$ & $j$ & $k_{\mathcal J}$ & $a^{k_{\mathcal J}}$ \\
\hline

$1s2s$ & ${}^3S$ & 1 & $\frac{1}{2}$ & 2 & 4.789 \\

\hline

$1s2p$ & ${}^3P$ 
      & 2 & $\frac{3}{2}$ & 2 & 5.204 \\
     & & 2 & $\frac{3}{2}$ & 4 & 4.927 \\
     & & 1 & $\frac{1}{2},\,\frac{3}{2}$ & 2 & 7.751 \\

$1s2p$ & ${}^1P$
     & 1 & $\frac{1}{2},\,\frac{3}{2}$ & 2 & 8.511 \\

\hline

$1s3s$ & ${}^3S$ 
     & 1 & $\frac{1}{2}$ & 2 & 19.324 \\

\hline

$1s3p$ & ${}^3P$ 
     & 2 & $\frac{3}{2}$ & 2 & 15.765 \\
     & & 2 & $\frac{3}{2}$ & 4 & 14.898 \\
     & & 1 & $\frac{1}{2},\,\frac{3}{2}$ & 2 & 22.622 \\

\hline

$1s3d$ & ${}^3D$
      & 3 & $\frac{5}{2}$ & 2 & 11.506 \\
     & & 3 & $\frac{5}{2}$ & 4 & 10.138 \\
     & & 3 & $\frac{5}{2}$ & 6 & 10.833 \\
     & & 2 & $\frac{3}{2},\,\frac{5}{2}$ & 2 & 19.511 \\
     & & 2 & $\frac{3}{2},\,\frac{5}{2}$ & 4 & 16.575 \\
     & & 1 & $\frac{3}{2}$ & 2 & 7.346 \\

$1s3d$ & ${}^1D$
      & 2 & $\frac{3}{2},\,\frac{5}{2}$ & 2 & 19.520 \\
     & & 2 & $\frac{3}{2},\,\frac{5}{2}$ & 4 & 16.582 \\

\hline

$1s3p$ & ${}^1P$
     & 1 & $\frac{1}{2},\,\frac{3}{2}$ & 2 & 24.193 \\

\hline
\end{tabular}
}
\end{table}

\begin{table}[htbp]
\centering
\scriptsize
\renewcommand{\arraystretch}{1.3}
\caption{Computed He\,\textsc{i} non-zero multi-level population-transfer
and polarization-transfer rates at $T=5000\,{\rm K}$. The coefficients
$a^k$ are defined through
$D^k(\alpha \mathcal{J}\rightarrow \alpha \mathcal{J}')
=a^k \times 10^{-9} n_{\rm H}$ s$^{-1}$, where $n_{\rm H}$ is in
cm$^{-3}$. Rates with $k=0$ correspond to population-transfer, while
rates with $k>0$ correspond to polarization-transfer. The listed channels
are ordered from the lower-energy to the higher-energy fine-structure
level according to the energies in Table~\ref{tab:nstar}. Reverse rates
are obtained from detailed balance. The columns preceding the tensorial
rank $k_{\mathcal J}$ list the valence-electron channels $j$ and $j'$ of
the simple-atom rate entering the recoupling.}
\label{tab:helium_multilevel_transfer_rates}
\resizebox{0.95 \columnwidth}{!}{%
\begin{tabular}{llcccccc}
\hline
Configuration & Term & $\mathcal{J}$ & $\mathcal{J}'$ & $j$ & $j'$ & $k_{\mathcal J}$ & $a^{k_{\mathcal J}}$ \\
\hline

$1s2p$ & ${}^3P$
      & 2 & 1 & $\frac{3}{2}$ & $\frac{1}{2}$ & 0 & 2.563 \\
     & & 2 & 1 & $\frac{3}{2}$ & $\frac{1}{2}$ & 2 & 0.391 \\
     & & 2 & 0 & $\frac{3}{2}$ & $\frac{1}{2}$ & 0 & 0.366 \\
     & & 1 & 0 & $\frac{3}{2}$ & $\frac{1}{2}$ & 0 & 0.801 \\

\hline

$1s3p$ & ${}^3P$
      & 2 & 1 & $\frac{3}{2}$ & $\frac{1}{2}$ & 0 & 7.879 \\
     & & 2 & 1 & $\frac{3}{2}$ & $\frac{1}{2}$ & 2 & 1.208 \\
     & & 2 & 0 & $\frac{3}{2}$ & $\frac{1}{2}$ & 0 & 1.123 \\
     & & 1 & 0 & $\frac{3}{2}$ & $\frac{1}{2}$ & 0 & 2.465 \\

\hline

$1s3d$ & ${}^3D$
      & 3 & 2 & $\frac{5}{2}$ & $\frac{3}{2}$ & 0 & 5.784 \\
     & & 3 & 2 & $\frac{5}{2}$ & $\frac{3}{2}$ & 2 & 9.554 \\
     & & 3 & 2 & $\frac{5}{2}$ & $\frac{3}{2}$ & 4 & -0.281 \\
     & & 3 & 1 & $\frac{5}{2}$ & $\frac{3}{2}$ & 0 & 3.028 \\
     & & 3 & 1 & $\frac{5}{2}$ & $\frac{3}{2}$ & 2 & 3.985 \\
     & & 2 & 1 & $\frac{5}{2}$ & $\frac{3}{2}$ & 0 & 2.980 \\
     & & 2 & 1 & $\frac{5}{2}$ & $\frac{3}{2}$ & 2 & 4.221 \\

\hline
\end{tabular}
}
\end{table}%
It is useful to note that if \(\mathcal{J}\ne \mathcal{J}'\),
the value \(k_\mathcal{J}=0\) is not allowed by the angular-momentum condition
\[
|\mathcal{J}-\mathcal{J}'|\leq k_\mathcal{J} \leq \mathcal{J}+\mathcal{J}' .
\]
Therefore, \(D^0(\alpha \mathcal{J} \mathcal{J}')=0\) for \(\mathcal{J}\ne \mathcal{J}'\). For this reason, only
the non-zero even tensorial ranks \(k_\mathcal{J}>0\) are listed for the relaxation
rates \(D^{k_\mathcal{J}}(\alpha \mathcal{J} \mathcal{J}')\). 
 In contrast, the rank \(k_\mathcal{J}=0\) transfer rates are not zero in general.
In fact, the
rank \(k_\mathcal{J}=0\) is allowed for diagonal pairs \((\mathcal{J},\mathcal{J})\) and
\((\mathcal{J}'',\mathcal{J}'')\); it is not allowed for off-diagonal coherences with
\(\mathcal{J}\ne \mathcal{J}'\) and/or \(\mathcal{J}''\ne \mathcal{J}'''\).

\begin{table}[htbp]
\centering
\scriptsize
\renewcommand{\arraystretch}{1.3}
\caption{He\,\textsc{i} non-zero multi-term depolarization coefficients $a^k$ defined through
$D^k(\alpha \mathcal{J} \mathcal{J}')=a^k\times 10^{-9} n_{\rm H}$ s$^{-1}$, where
$n_{\rm H}$ is in cm$^{-3}$. Only even tensorial ranks are listed. The column preceding the tensorial rank $k_{\mathcal J}$ lists the valence-electron channel(s) $j$ of the simple-atom rate entering the recoupling. The multi-term rates $D^k(\alpha \mathcal{J} \mathcal{J}') \!=\! D^k(\alpha \mathcal{J}' \mathcal{J})$ are built from the simple-atom depolarization rates $D^{k_j}(j)$. When more than one channel contributes, all contributing channels are listed and their partial contributions are summed.}
\label{tab:helium_multiterm_rates}
\resizebox{0.95 \columnwidth}{!}{%
\begin{tabular}{llccc@{\hspace{8pt}}c}
\hline
Configuration & Term & $(\mathcal{J},\mathcal{J}')$ & $j$ & $k_{\mathcal J}$ & $a^{k_{\mathcal J}}$ \\
\hline

$1s2s$ & ${}^3S$
      & $(1,1)$ & $\frac{1}{2}$ & 2 & 4.789 \\

\hline

$1s2p$ & ${}^3P^\circ$
      & $(2,2)$ & $\frac{3}{2}$ & 2 & 5.204 \\
     & & $(2,2)$ & $\frac{3}{2}$ & 4 & 4.927 \\
     & & $(2,1)$ & $\frac{3}{2}$ & 2 & 6.357 \\
     & & $(1,1)$ & $\frac{1}{2},\,\frac{3}{2}$ & 2 & 7.751 \\

$1s2p$ & ${}^1P^\circ$
      & $(1,1)$ & $\frac{1}{2},\,\frac{3}{2}$ & 2 & 8.511 \\

\hline

$1s3s$ & ${}^3S$
      & $(1,1)$ & $\frac{1}{2}$ & 2 & 19.324 \\

\hline

$1s3p$ & ${}^3P^\circ$
      & $(2,2)$ & $\frac{3}{2}$ & 2 & 15.765 \\
     & & $(2,2)$ & $\frac{3}{2}$ & 4 & 14.898 \\
     & & $(2,1)$ & $\frac{3}{2}$ & 2 & 19.644 \\
     & & $(1,1)$ & $\frac{1}{2},\,\frac{3}{2}$ & 2 & 22.622 \\

\hline

$1s3d$ & ${}^3D$
      & $(3,3)$ & $\frac{5}{2}$ & 2 & 11.506 \\
     & & $(3,3)$ & $\frac{5}{2}$ & 4 & 10.138 \\
     & & $(3,3)$ & $\frac{5}{2}$ & 6 & 10.833 \\
     & & $(3,2)$ & $\frac{5}{2}$ & 2 & 12.623 \\
     & & $(3,2)$ & $\frac{5}{2}$ & 4 & 10.146 \\
     & & $(2,2)$ & $\frac{3}{2},\,\frac{5}{2}$ & 2 & 19.511 \\
     & & $(2,2)$ & $\frac{3}{2},\,\frac{5}{2}$ & 4 & 16.575 \\
     & & $(2,1)$ & $\frac{3}{2}$ & 2 & 10.822 \\
     & & $(1,1)$ & $\frac{3}{2}$ & 2 & 7.346 \\

$1s3d$ & ${}^1D$
      & $(2,2)$ & $\frac{3}{2},\,\frac{5}{2}$ & 2 & 19.520 \\
     & & $(2,2)$ & $\frac{3}{2},\,\frac{5}{2}$ & 4 & 16.582 \\

\hline

$1s3p$ & ${}^1P^\circ$
      & $(1,1)$ & $\frac{1}{2},\,\frac{3}{2}$ & 2 & 24.193 \\

\hline
\end{tabular}
}
\end{table}

\begin{table*}[htbp]
\centering
\scriptsize
\renewcommand{\arraystretch}{1.3}
\caption{Computed He\,\textsc{i} non-zero multi-term  collisional transfer
rates at $T=5000\,{\rm K}$. The values $a^{k_\mathcal{J}}$ are defined through
$D^k(\alpha \mathcal{J} \mathcal{J}' \rightarrow \alpha \mathcal{J}''\mathcal{J}''') =
a^{k_\mathcal{J}} \times 10^{-9} n_{\rm H}$ s$^{-1}$, where $n_{\rm H}$ is in cm$^{-3}$.
Only even tensorial ranks $k_{\mathcal J}$ are retained. The rank $k_\mathcal{J}=0$ cases correspond to
population-transfer rates between diagonal pairs, while the cases with
$k_\mathcal{J}>0$ correspond to polarization-transfer rates. The columns preceding the tensorial rank $k_{\mathcal J}$ give the valence-electron channels $j$ and $j'$ of the simple-atom transfer rate $D^{k_j}(j\rightarrow j')$ entering the recoupling; in the present He\,\textsc{i} model each multi-term transfer reduces to a single off-diagonal channel, $j\neq j'$.}
\label{tab:helium_multiterm_pair_transfer}
\resizebox{0.95 \columnwidth}{!}{%
\begin{tabular}{llcccc@{\hspace{8pt}}c@{\hspace{8pt}}c}
\hline
Configuration & Term & $(\mathcal{J},\mathcal{J}')$ & $(\mathcal{J}'',\mathcal{J}''')$ & $j$ & $j'$ & $k_{\mathcal J}$ & $a^{k_{\mathcal J}}$ \\
\hline

$1s2p$ & ${}^3P^\circ$ 
      & $(1,1)$ & $(2,2)$ & $\frac{1}{2}$ & $\frac{3}{2}$ & 0 & 4.272 \\
    & & $(1,1)$ & $(2,2)$ & $\frac{1}{2}$ & $\frac{3}{2}$ & 2 & 0.651 \\
    & & $(1,1)$ & $(1,2)$ & $\frac{1}{2}$ & $\frac{3}{2}$ & 2 & 0.426 \\
    & & $(1,1)$ & $(2,1)$ & $\frac{1}{2}$ & $\frac{3}{2}$ & 2 & -0.426 \\
    & & $(0,0)$ & $(2,2)$ & $\frac{1}{2}$ & $\frac{3}{2}$ & 0 & 1.831 \\
    & & $(0,0)$ & $(1,1)$ & $\frac{1}{2}$ & $\frac{3}{2}$ & 0 & 2.403 \\

\hline
$1s3p$ & ${}^3P^\circ$
      & $(1,1)$ & $(2,2)$ & $\frac{1}{2}$ & $\frac{3}{2}$ & 0 & 13.131 \\ 
    & & $(1,1)$ & $(2,2)$ & $\frac{1}{2}$ & $\frac{3}{2}$ & 2 & 2.013 \\
    & & $(1,1)$ & $(1,2)$ & $\frac{1}{2}$ & $\frac{3}{2}$ & 2 & 1.318 \\
    & & $(1,1)$ & $(2,1)$ & $\frac{1}{2}$ & $\frac{3}{2}$ & 2 & -1.318 \\
    & & $(0,0)$ & $(2,2)$ & $\frac{1}{2}$ & $\frac{3}{2}$ & 0 & 5.617 \\
    & & $(0,0)$ & $(1,1)$ & $\frac{1}{2}$ & $\frac{3}{2}$ & 0 & 7.395 \\

\hline
$1s3d$ & ${}^3D$ 
      & $(2,2)$ & $(3,3)$ & $\frac{3}{2}$ & $\frac{5}{2}$ & 0 & 8.097 \\
    & & $(2,2)$ & $(3,3)$ & $\frac{3}{2}$ & $\frac{5}{2}$ & 2 & 13.375 \\
    & & $(2,2)$ & $(3,3)$ & $\frac{3}{2}$ & $\frac{5}{2}$ & 4 & -0.394 \\

    & & $(2,2)$ & $(3,2)$ & $\frac{3}{2}$ & $\frac{5}{2}$ & 2 & 5.981 \\
    & & $(2,2)$ & $(3,2)$ & $\frac{3}{2}$ & $\frac{5}{2}$ & 4 & 0.188 \\            
    & & $(2,2)$ & $(2,3)$ & $\frac{3}{2}$ & $\frac{5}{2}$ & 2 & -5.981 \\    
    & & $(2,2)$ & $(2,3)$ & $\frac{3}{2}$ & $\frac{5}{2}$ & 4 & -0.188 \\

    & & $(2,1)$ & $(3,3)$ & $\frac{3}{2}$ & $\frac{5}{2}$ & 2 & 9.313 \\
    & & $(1,2)$ & $(3,3)$ & $\frac{3}{2}$ & $\frac{5}{2}$ & 2 & -9.313 \\

    & & $(2,1)$ & $(2,3)$ & $\frac{3}{2}$ & $\frac{5}{2}$ & 2 & 11.397 \\
    & & $(1,2)$ & $(3,2)$ & $\frac{3}{2}$ & $\frac{5}{2}$ & 2 & 11.397 \\
    & & $(2,1)$ & $(3,2)$ & $\frac{3}{2}$ & $\frac{5}{2}$ & 2 & 18.214 \\
    & & $(1,2)$ & $(2,3)$ & $\frac{3}{2}$ & $\frac{5}{2}$ & 2 & 18.214 \\

    & & $(1,1)$ & $(3,3)$ & $\frac{3}{2}$ & $\frac{5}{2}$ & 0 & 7.065 \\
    & & $(1,1)$ & $(3,3)$ & $\frac{3}{2}$ & $\frac{5}{2}$ & 2 & 9.298 \\

    & & $(1,1)$ & $(3,2)$ & $\frac{3}{2}$ & $\frac{5}{2}$ & 2 & 3.786 \\
    & & $(1,1)$ & $(2,3)$ & $\frac{3}{2}$ & $\frac{5}{2}$ & 2 & -3.786 \\

    & & $(2,1)$ & $(2,2)$ & $\frac{3}{2}$ & $\frac{5}{2}$ & 2 & 7.393 \\
    & & $(1,2)$ & $(2,2)$ & $\frac{3}{2}$ & $\frac{5}{2}$ & 2 & -7.393 \\

    & & $(1,1)$ & $(2,2)$ & $\frac{3}{2}$ & $\frac{5}{2}$ & 0 & 4.966 \\
    & & $(1,1)$ & $(2,2)$ & $\frac{3}{2}$ & $\frac{5}{2}$ & 2 & 7.035 \\

\hline
\end{tabular}
}
\end{table*}

\section{Conclusion}

We have presented a new set of collisional depolarization, polarization-transfer, and population-transfer rates, due to isotropic collisions with neutral hydrogen, for the singlet and triplet terms of neutral helium that control the main solar He\,\textsc{i} diagnostics, including the $10830\,\text{\AA}$ and D$_3$ multiplets. The rates have been computed in both the multi-level and multi-term formulations and are tabulated in Tables~\ref{tab:helium_multilevel_depolarization_rates}, \ref{tab:helium_multilevel_transfer_rates}, \ref{tab:helium_multiterm_rates}, and \ref{tab:helium_multiterm_pair_transfer} for $T = 5000\,\mathrm{K}$, together with scaling laws for chromospheric and prominence temperatures. 
 These results address an important missing ingredient in the modeling of He\,\textsc{i} spectropolarimetry. The $10830\,\text{\AA}$ and $5876\,\text{\AA}$ lines are among the most sensitive probes of magnetic fields in the upper solar atmosphere, where their polarization is shaped by atomic level polarization, the Hanle effect, and the Zeeman effect. Forward-modeling and inversion numerical codes solve multi-level or multi-term SEE for the He\,\textsc{i} density matrix; the rates given here can therefore be incorporated directly, replacing the order-of-magnitude estimates commonly adopted so far. 

Our results provide a quantitative basis for reassessing the role of neutral-hydrogen collisions in destroying or transferring atomic polarization. In particular, they make it possible to test the assumption usually adopted in solar physics  that collisional depolarization is generally expected to be negligible, especially in typical quiescent prominences, because 
the required hydrogen densities are too high. At the same time, our tabulated rates should allow one to identify denser regimes---such as active-region filaments, dense spicules, post-flare loops, and coronal rain---where collisional depolarization rates may become comparable to radiative rates. They also separate the effects of transfer rates from depolarization rates, which are often merged in simplified treatments.

\begin{acknowledgements}
This research work was funded by Institutional Fund
Projects under grant no. (IFPIP:1001-130-1443). The authors gratefully acknowledge technical and financial support provided by the Ministry of Education and
King Abdulaziz University, DSR, Jeddah, Saudi Arabia
\end{acknowledgements}

\end{document}